\documentclass[fleqn,usenatbib]{mnras}

\usepackage{newtxtext,newtxmath}

\usepackage[T1]{fontenc}

\DeclareRobustCommand{\VAN}[3]{#2}
\let\VANthebibliography\thebibliography
\def\thebibliography{\DeclareRobustCommand{\VAN}[3]{##3}\VANthebibliography}

\usepackage{graphicx}	
\usepackage{amsmath}	
\DeclareSymbolFont{matha}{OML}{txmi}{m}{it}
\DeclareMathSymbol{\varv}{\mathord}{matha}{118}

\title[Slow magnetoacoustic waves in stratified two-fluid plasmas]{Slow magnetoacoustic waves in gravitationally stratified two-fluid plasmas in strongly ionised limit}

\author[A. Alharbi et al.]{
A. Alharbi,$^{1,2}$\thanks{E-mail: aalharbi8@sheffield.ac.uk}
I. Ballai,$^{1}$
V. Fedun$^{3}$
and G. Verth$^{1}$
\\
$^{1}$Plasma Dynamics Group, School of Mathematics and Statistics, University of Sheffield, Housnfield Road, Hicks Building, Sheffield, S3 7RH, UK\\
$^{2}$Department of Mathematics, Jamoum University College, Umm Al-Qura University, Jamoum, Makkah, Saudi Arabia\\
$^{3}$Plasma Dynamics Group, Department of Automatic Control and Systems Engineering, The University of Sheffield, Mappin Street, Sheffield, S1 3JD, UK
}

\date{Accepted XXX. Received YYY; in original form ZZZ}

\pubyear{2020}

\begin{document}
\label{firstpage}
\pagerange{\pageref{firstpage}--\pageref{lastpage}}
\maketitle
\begin{abstract}
The plasma dynamics at frequencies comparable with collisional frequency between various species has to be described in multi-fluid framework, where  collisional interaction between particles is an important ingredient. In our study we will assume that charged particles are strongly coupled, meaning that they form a single fluid that interacts with neutrals, therefore we will employ a two-fluid model.  
   Here we aim to investigate the evolutionary equation of slow sausage waves propagating in a gravitationally stratified flux tube in the two-fluid solar atmosphere in a strongly ionised limit using an initial value analysis.
   Due to the collisional interaction between massive particles (ions and neutrals) the governing equations are coupled. Solutions are sought in the strongly ionised limit and the density ratio between neutrals and charged particles is a small parameter. This limit is relevant to the upper part of the chromosphere.
   Our results show that slow sausage waves associated with charged particles propagate such that their possible frequency is affected by a cut-off due to the gravitational stratification. In contrast, for neutral acoustic waves the cut-off value  applies on their wavelength and only small wavelength waves are able to propagate.  Slow modes associated to neutrals are driven by the collisional coupling with ions.
\end{abstract}




\section{Introduction}
One of the key characteristics of solar atmosphere is that in the lower regions (photosphere and chromosphere) the plasma is partially ionised, where neutral atoms, electrons and positively charged ions can interact through short and long-range collisions. The ionisation degree of the plasma depends mainly on temperature. Heinzel et al. (2015) showed that in the case solar prominences, the ionisation degree also depends on density and pressure. Solar atmospheric models such as the VAL (Vernazza et al. 1981) or FAL (Fontenla et al. 1990) models predict a very low ionisation degree in the deep photosphere (where for every ion there are approximately $10^4$ neutrals), and increases with height due to the increase of temperature. The different species of particles present in the plasma  interact through collisions and the frequency of the collisions also decreases with height  due to the decrease of density of particles with height (a quantitative description is presented in Section 2.).  Although collisions between various species are important for various aspects related to partially ionised plasmas such as thermalisation of the plasma, various ionisation/recombination processes, appearance of thermal layers for shock waves in partially ionised plasmas, etc. (Shanmugasundaram and Murty 1978, Mathers 1980, Terradas et al. 2015, Mart\'inez-G\'omez et al. 2018, Ballai 2019, Kuzma et al. 2020), the short-range collisions between neutrals and charged particles are important as only this physical mechanism ensures that neutrals are a constituent part of the plasma. Given the large mass difference between electrons and ions, effective way for transferring energy and momentum occurs only via collisions between ions and neutrals.  The collisions between electrons and the ions/neutrals help in the Maxwellisation of the electron population but is not affecting considerably the energy and momentum of massive particles. 

Unfortunately the characteristics of current ground-based and space-borne observational facilities are not suitable for the direct observation of waves with frequencies of the order of the collisional frequencies in partially ionised plasmas, as these waves require a time cadence that currently cannot be achieved. Nevertheless, waves and instabilities in partially ionised plasmas together with their effects have been largely explored theoretically and numerically. For a recent review on the progress of research on dynamical processes in solar and astrophysical plasmas see Ballester et al. (2018).  

The theoretical investigation of waves in partially ionised plasmas under solar conditions has received recently an increased attention. More and more studies started to take into account the realistic model of a solar atmosphere, where the plasma is not hot enough to ensure a full ionisation. We should mention here that the consideration of partial ionisation effects depend on the range of frequencies we are interested in. If the frequency of waves we plan to investigate is much smaller than the collisional frequency of particles, the dynamics can be described within the framework of magnetohydrodynamics (MHD). Since current observational capabilities are mostly centered onto this regime, the observation of waves outside the MHD regime can be achieved only indirectly. Although the observation of waves is still nearly impossible, several attempts have been made to evidence the effect of partial ionisation in solar lower atmosphere.  Due to an imperfect collisional coupling between massive particles (ions and neutrals), there is an imbalance in the velocity of these species and this has been evidenced through a simultaneous measuring of the Doppler shift in the Fe II ion and neutral Fe I lines over the same volume of plasma in the sunspot penumbra (Khomenko et al. 2015). Later, Khomenko et al. (2016) found non-negligible differences in He I and Ca II velocities in solar prominences. Gilbert et al. (2007) compared He I and H$\alpha$ data in multiple solar prominences in different phases of their life cycle and detected the drainage effect across the prominence magnetic field with different timescales for He and H atom. Later, de la Cruz Rodrigez and Socas-Navarro (2011) have reported misalignment in the visible direction of chromospheric fibrils that were attributed to the large ambipolar diffusion, that is, when the ion-neutral collisional frequency drops, the magnetic field can slip through the neutral population. This observational result has been later confirmed through numerical simulations by Mart\'inez-Sykora et al. (2016) using  advanced  radiative MHD simulations. Finally, some observations have found evidence for higher frequency waves with periods as short as 45 s (0.15 Hz) in spicules (Okamoto and De Pontieu 2011). Transition region spectral lines often show significant broadening beyond the thermal width of the order of 20 km s$^{-1}$ in exposure times as short as 4 s (De Pontieu et al. 2015). If this non-thermal broadening were to be
caused by waves, wave frequencies could be significantly higher than 1 Hz.

The framework used to describe the dynamics of waves in partially ionised plasmas depends on the frequency range of interest. For wave frequencies that are of the order of the collisional frequency between ions and neutrals we can employ a model where charged and neutrals particles are treated as separate, but interacting, fluids. Waves propagating in partially ionised plasmas differ qualitatively and quantitatively from their counterpart in fully ionised plasmas. First of all, the spectrum of possible waves is larger as now, in addition to the ion-related modes, there are also acoustic modes associated to neutrals.  

The study of waves in inhomogeneous plasma is not an easy task as inhomogeneities can change dramatically the property of waves. The damping of Alfv\'en waves in gravitationally stratified plasmas and their contribution to the heating of chromospheric plasma has been studied by a number of authors. Leake et al. (2005) used a single-fluid plasma approximation in the presence of Cowling resistivity and they found a very clear frequency-dependent damping of waves for chromospheric heights of 1000-2500 Km above the solar surface. According to these authors Alfv\'en waves with frequencies below 0.01 Hz are unaffected by dissipative effects and propagate through the partially
ionised plasma with little diffusion. In contrast, Alfv\'en waves with frequency above 0.6 Hz are completely damped. The research in this topic has been extended later by Tu and Song (2013), who carried out a numerical investigation of the two-fluid approximation, where collisions between various species (neutrals, electrons and positive ions) have been considered. The results of this analysis show that thanks to the density gradient, Alfv\'en waves are partially reflected throughout the chromosphere and more strongly at higher altitudes. Waves were observed to be damped in the lower chromosphere dominantly through Joule dissipation, producing heating strong enough to balance the radiative loss for the quiet chromosphere without invoking anomalous processes or turbulences. These authors also found that there is an upper cutoff frequency, depending on the background magnetic field, above which waves are completely damped. For a magnetic field of 100 G, the determined cut-off (or critical) frequency was found to be 0.12 Hz. On the other hand the damping of Alfv\'en waves can also be used to infer key plasma parameters by means of seismological techniques. For example, recent observations by Kohutova et al. (2020) showed torsional Alfv\'en waves propagating with a period of 89 s, an amplitude of 41 km s$^{-1}$, and a damping time of 136 s. Using a single-fluid partially ionised plasma model for prominences, Ballai (2020) employed the observations by Kohutova et al. (2020) in conjunction with the dispersion relation for torsional Alfv\'en waves to diagnose the ionisation degree of the plasma, and found that the neutral number density of the plasma was $5.08\times 10^{16}$ m$^{-3}$. A numerical analysis of the variation of the cut-off frequency with height has recently been made by W\'ojcik et al. (2019) assuming a two-fluid plasma.

The properties of magnetoacoustic waves propagating in a two-fluid homogeneous plasmas has been studied by a number of authors. Zaqarashvili et al. (2011) and later Soler et al. (2013) have shown that the collisional frequency between ions and neutrals can considerably modify the behaviour and properties of waves. When applied to chromospheric situations the study by these authors revealed that wavelengths smaller than $10^3$ m are affected by two-fluid effects in the presence of strong magnetic fields. However, their approach is an eigenvalue problem, meaning that the temporal evolution of waves cannot be studied. Furthermore, they neglected gravitational stratification, meaning that important effects such as the presence of frequency cut-offs could not be studied. In the present investigation we plan to address both of these shortcomings.

The paper is structured as follows: The physical assumptions and the mathematical background necessary to carry out our research is presented in Section 2. The evolutionary equations describing the spatial and temporal evolution of slow sausage modes attached to each species will be derived in Section 3. Solutions of these equations assuming a strong ionisation thermal equilibrium is obtained in Section 4. The asymptotic solutions corresponding to large values of time is obtained in Section 5 assuming a sinusoidal pulse driver for typical solar chromospheric conditions. Finally, our results are summarised and discussed in Section 6.



\section{Assumptions and mathematical background}

Before we embark on describing the evolution of slow guided waves in a gravitationally stratified plasma we need to make a few assumptions that will simplify our analysis. First of all we assume that during the typical time-scales involved in wave description the plasma remains in ionisation equilibrium, i.e. no additional ions are created by ionisation or neutrals due to recombination. This assumption is rather restrictive as typical time-scales associated to ionisation and recombination often can be comparable to period of waves. A treatment of waves in partially ionised non-equilibrium plasma can be found in the study by Ballai (2019).

Waves will propagate in a vertically unbounded magnetic cylinder and the magnetic field is parallel to the symmetry axis of the flux tube in the positive $z$-direction. The environment of the flux tube is non-magnetic. 
 To simplify our analysis we will assume that the flux tube of cross-sectional area $A(z,t)$ is thin, i.e. waves propagating in the flux tube have wavelengths much larger than the radius of the tube (also known as the slender tube approximation). In this limit waves will not "sense" the boundary of waveguide. Due to the gravitational stratification, the dispersion relation of slow waves becomes $\omega^2\approx k^2c_T^2+\omega_c^2$ (Roberts and Webb 1978), where $k$ is wave number, $c_T$ is cusp speed and $\omega_c$ is the cut-off frequency of waves that depends on characteristic speeds and gravitational acceleration. As a result, the frequency of waves is increased compared to the unstratified case and waves become dispersive, which means that waves with longer wavelength will propagate faster. In the opposite case, when the wavelength of waves is comparable (or smaller) to the radius of the flux tube we are dealing with a thick flux tube, where the properties of waves are considerably changed. The propagation characteristics of slow sausage waves in a thick flux tube in a fully ionised plasma has been investigated by Pardi et al. (2014). 

Since we aim to study the propagation of longitudinal waves, we can consider only the longitudinal velocity components of the species involved in the problem. 

We assume that the length scales of variables that describe the dynamical and thermodynamical state of the plasma are much longer than the scattering mean free path, so that the concept of fluid is applicable. We are going to employ a two-fluid approximation where neutrals and charged species will interact through collisions. Although the interaction between electrons and neutrals also takes place, we are going to limit ourselves to the collisions between the massive particles, i.e. ions and neutrals. For simplicity we are going to label the charged species as "ions". Physical quantities related to this fluid are labeled by an index {\it i} and the parameters of the neutral fluid will be labelled by an index {\it n}. We should mention here that the charged particles (ions and electrons) all have a common velocity since differences in the divergence of the ion and electron velocities would lead to charge separation and strong electric fields opposing the charge separation. see .

The system of equations describing the linear dynamics of the two-fluid plasma  (for details see, Khomenko et al. 2014) is given by 
\begin{equation} \label{eq1}
     \frac{\partial}{\partial t}(\rho_{0i} A+\rho_i A_0)+\frac{\partial}{\partial z}(\rho_{0i} A_0 \varv_{i})=0,
 \end{equation}
  \begin{equation} \label{eq2}
     \frac{\partial}{\partial t}(\rho_{0n} A+\rho_n A_0)+\frac{\partial}{\partial z}(\rho_{0n} A_0 \varv_{n} )=0,
 \end{equation}
\begin{equation}\label{eq3}
    \rho_{0i} \frac{\partial \varv_{i}}{\partial t}+ \frac{\partial p_i}{\partial z}+ \rho_i g+\alpha_{in}(\varv_{i}-\varv_{n})=0,
\end{equation}
\begin{equation}\label{eq4}
    \rho_{0n} \frac{\partial \varv_{n}}{\partial t}+ \frac{\partial p_n}{\partial z}+ \rho_n g+\alpha_{in}(\varv_{n}-\varv_{i})=0,
\end{equation}
\begin{equation}  \label{eq5}
    \frac{\partial p_i}{\partial t}+ \varv_{i} \frac{dp_{0i}}{dz}=c_{Si}^2 \left(\frac{\partial \rho_i}{\partial t}+\varv_{i} \frac{\partial \rho_{0i}}{\partial z} \right),
\end{equation}
\begin{equation}  \label{eq6}
    \frac{\partial p_n}{\partial t}+ \varv_{n} \frac{dp_{0n}}{dz}=c_{Sn}^2 \left(\frac{\partial \rho_n}{\partial t}+\varv_{n}\frac{\partial \rho_{0n}}{\partial z} \right).
\end{equation}
 Mathematical details of the governing equations can be found in earlier studies by Defouw (1976) and Herbold et al. (1985). The above system of equations has to be supplemented by the two conditions
\begin{equation}\label{con}
    B_0A+BA_0=0, \quad p_i+p_n+\frac{B_0}{\mu_0}B=\pi(z,t)
\end{equation}
expressing the conservation (in a linearised way) of the magnetic flux, and the total pressure at the boundaries of the flux tube. The quantities with an index '0' denote equilibrium values. In the above equations $\rho_i$, $v_i$ and $p_i$ are the density, longitudinal velocity component and pressure of charged particles (ions and electrons), $\rho_n$, $v_n$ and $p_n$ are the corresponding quantities for neutral species, $g=274$ $ m/s^2$ the constant gravitational acceleration, $c_{Si}=\left(\gamma {p_{0i}}/\rho_{0i}\right)^{1/2}$ is the ion sound speed and $c_{Sn}=\left(\gamma {p_{0n}}/\rho_{0n}\right)^{1/2}$ is the neutral sound speed and $\gamma$ is ratio of specific heats that will be considered constant ($\gamma=5/3$).  In Equation (\ref{con}) $B_0$ and $B$ are the equilibrium and perturbed magnetic field, $A_0$ and $A$ are the equilibrium cross-section area of the tube and the associated perturbation, while in the pressure balance equation $\pi(z,t)$ is the external pressure, and $\mu_0$ is the permeability of free space. We should mention that, strictly speaking, the energy conservation equations for the two species should have contained a term that describe the energy lost due to the collisional friction between particles, however, since this term is nonlinear (proportional to the square of ($\varv_{i}-\varv_{n}$), these will be neglected and the energy conservation is described by an adiabatic equation  written for each fluid.

During the propagation of waves particles will undergo collisions with other particles. Neglecting mutual collisions between particles of the same type, the frictional coefficients between the colliding ions and neutrals is $\alpha_{in}$ and is given by
\begin{equation}
\alpha_{in}=\rho_i \nu_{in}=\rho_n \nu_{ni},
\label{eq6.1}
\end{equation}
where $\nu_{in}$ and $\nu_{ni}$ are the frequencies of ion-neutral and neutral-ions collisions. In the above equation the frictional coefficient between ions and neutrals is given by (Braginskii 1965)
\begin{equation}
\alpha_{in}=2n_in_nm_i\sigma_{in}\left(\frac{k_BT}{\pi m_i}\right)^{1/2},
\label{eq6.2}
\end{equation}
where $\sigma_{in}=5\times 10^{-19}$ m$^2$ is the collisional cross-section (Vranjes and Krstic 2013), $k_B$ is the Boltzmann constant, $m_i$ is the ion mass and $n_i$ and $n_n$ are the number density of ions and neutrals, respectively. In the above calculations we assumed that we are dealing with hydrogen plasma. Although normally the collisional frequencies are also height dependent, we are going to treat these quantities as constants and we are going to evaluate them for particular solar parameters, at particular height. 

The collisions between the massive particles in the system acts as a dissipative term and waves will be expected to decay due to the collisions. Using the standard atmospheric models it can be shown that up to a height of approximately 2 Mm $\nu_{in}>\nu_{ni}$, however, after this height, this inequality reverses due to the decrease in the number of neutrals thanks to the ionisation driven by the increase in temperature.

Due to the gravitational stratification equilibrium quantities will have a height-dependence. In a hydrostatic equilibrium the variation of the pressure with height for the two constituent fluids is given by
\[
p_{0i}(z)=p_{0i}(0)e^{-\gamma_i(z)}, \quad  p_{0n}(z)=p_{0n}(0)e^{-\gamma_n(z)}.
\]
The dimensionless quantities $\gamma_i(z)$ and $\gamma_n(z)$ are given by
\[
\gamma_i(z)=\int_0^z\frac{1}{H_i(z')}dz', \quad \gamma_n(z)=\int_0^z\frac{1}{H_n(z')}dz',
\]
where 
\[
H_i(z)=\frac{RT_i(z)}{{\tilde \mu_i}g}, \quad H_n(z)=\frac{RT_n(z)}{{\tilde \mu_n}g}
\]
are the gravitational pressure scale heights for ions and neutrals, $R$ is the gas constant, ${\tilde \mu_i}$ and ${\tilde \mu_n}$ are the mean atomic weights and $T_i$ and $T_n$ are the temperature of the ion and neutral fluid (such that the mean translational kinetic energy or fluid particle in a frame moving with the fluid is $(3/2)k_BT_i$ and $(3/2)k_BT_n$, respectively). We should stress out that $T_i$ stands for the temperature of the charged fluid, therefore it is the sum of the temperatures corresponding to ions and electrons. For simplicity we assume that the plasma is isothermal,  meaning that the temperatures do not depend on height. As a result, the scale-heights are also constant, so the height-variation of the two pressures are simply given by
\[
p_{0i}=p_{0i}(0)e^{-z/H_i}, \quad p_{0n}=p_{0n}(0)e^{-z/H_n}.
\]
Using the ideal gas law the equilibrium densities of the two species also vary according to similar laws. One important implication of the isothermal limit is that the sound speeds for the two species will be constant and the two scale-heights will be simply $H_i=c_{Si}^2/\gamma g$ and $H_n=c_{Sn}^2/\gamma g$. 

 As a consequence of the variation of pressure and density with height, the equilibrium is reached if the magnetic field and the magnetic flux tube's cross section area vary with height according to (for explanation see, e,g, Roberts and Webb 1978)
\[
B_0(z)=B_0(0)e^{-z/2H_i}, \quad A_0(z)=A_0(0)e^{z/2H_i}.
\]
With this particular choice of the magnetic field even the Alfv\'en speed, defined as,
\[
v_A=\frac{B_0(z)}{(\mu_0\rho_{0i}(z))^{1/2}},
\]
becomes also height-independent. 

Our calculations will be further simplified by considering that temporal changes in the environment (the plasma outside the magnetic flux tube) take place over time scales that are much longer than any characteristic times scales of interest occurring inside the flux tube (very often this is also called the rigid boundary approximation). As a result every term that contains a time derivative of the external pressure, $\pi(z,t)$ will be neglected. 

The propagation of slow waves in an unbounded plasma has been investigated previously in great detail as an eigenvalue problem by Zaqarashvili et al. (2011) and Soler et al. (2013) for varying collisional rate between ions and neutrals. While in the collisionless limit the two slow modes propagate with the ion cusp speed, and neutral sound speed, respectively, in the collisional case the propagation speed of slow waves become complex due to their interaction. In the weakly ionised and very low plasma-beta regime these authors found that the neutral slow waves  are affected by a frequency cut-off, while the slow mode associated to ions becomes the modified slow mode
\[
\omega^2\approx k^2\frac{c_{Si}^2+\chi c_{Sn}^2}{1+\chi},
\]
where $\chi$ is defined as $\chi=\rho_{0n}/\rho_{0i}$. When $\chi\ll 1$, the propagation speed of ion slow waves becomes essentially $\omega^2\approx k^2c_{Si}^2$. 

Since we aim to analyze the spatial and temporal evolution of waves, we will not discuss explicitly the role of collisions as in the study by Soler et al. (2013), instead we will assume a fixed value of the collisional frequency that is representative for the region of the solar atmosphere where our analysis is valid. In our study we will also assume that the parameter $\chi$ is much less  than one and this parameter can be used as an expansion parameter to simplify the mathematics. Accordingly, the density ratio, $\chi$, can be written as
\begin{equation}
    \chi =\frac{\rho_{0n}(z)}{\rho_{0i}(z)} =\frac{\rho_{0n}(0)}{\rho_{0i}(0)} \exp\left[-z\left(\frac{1}{H_n}-\frac{1}{H_i}\right)\right]=\chi_0 e^{-z/h}.
    \label{eq10}
\end{equation}
Clearly the condition $\chi\ll 1$ means not only that $\chi_0\ll 1$, but also that $h>0$, i.e. $H_i>H_n$. This assumption is based on the relative variation of the neutral density compared to the density of ions with height according to the AL C7 atmospheric model (Avrett and Loeser 2008). In Fig 1 we compare the predictions of the VAL IIIC atmospheric model (Vernazza et al. 1981) shown by red line, with the AL C7 model shown by the blue line. Clearly the two models show a good similarity up to heights of about 2 Mm. The large discrepancy following this height is due to more extensive set of chromospheric observations. 
 \begin{figure}
   	\includegraphics[width=\columnwidth]{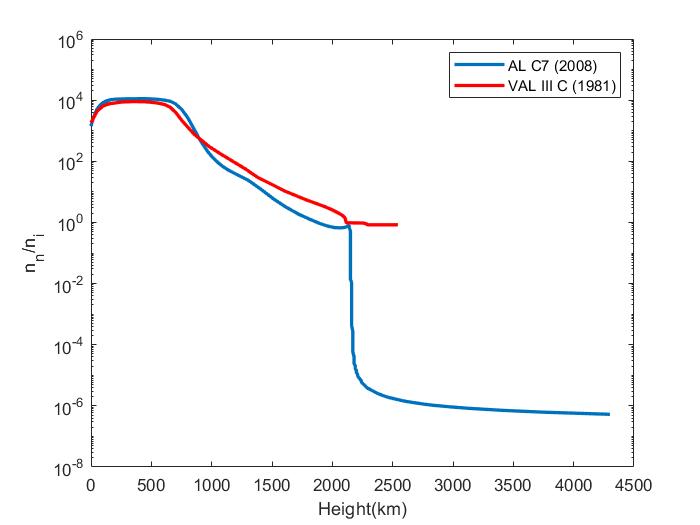}
    \caption{The variation of the ratio of number densities of neutrals and ions with height based on the VAL III C atmospheric model (Vernazza et al. 1981, red line) and the AL c7 atmospheric model (Avrett and Loeser (2008, blue line).}
   \end{figure}
 It is clear that, while in the photosphere the density ratio can be even of the order of $10^4$, for chromospheric heights the density ratio is very small, and, therefore, our assumption is justified.


\section{Evolutionary equations}
The governing Equations (\ref{eq1})--(\ref{eq6}) together with the particular choice of equilibrium parameters and the two conservation relations can be reduced to a system of coupled differential equations for the longitudinal velocity components of the two fluids of the form
\begin{equation}\label{eq11}
     \frac{\partial^2 v_i}{\partial t^2}-c_T^{2} \frac{\partial^2 v_i}{\partial z^2}+ \alpha_1 \frac{\partial v_i}{\partial z}+ \alpha_2 v_i =0,  
\end{equation}
\[
      \frac{\partial^2 v_n}{\partial t^2}-c_{Sn}^{2}  \frac{\partial^2 v_n}{\partial z^2}+\alpha_3 \frac{\partial v_n}{\partial z}+\alpha_4 v_n+\nu_{ni} \frac{\partial v_n}{\partial t}=
      \]
      \begin{equation}
      \label{eq12}
      -\frac{c_{Sn}^2 c_T^{2}}{v_A^{2}}\frac{\partial^2 v_i}{\partial z^2}+\alpha_5 \frac{\partial v_i}{\partial z}+\alpha_6 v_i,
\end{equation}
where $c_T^2=c_{Si}^{2} v_A^{2}/(c_{Si}^2+v_A^2)$ is the cusp speed related to ions. The above two relations describe the evolution of two slow magnetoacoustic modes (the ion-acoustic mode propagating with the cusp speed and neutral-acoustic mode propagating with the neutral sound speed). We should note that due to the relative low number of neutrals ion-acoustic modes will propagate (to leading order) unaffected by collisions, while the propagation of neutral-acoustic modes is strongly affected by collisions with ions and their dynamical behaviour is driven by ions through the set of terms on the right-hand side of Equation (\ref{eq12}).

The coefficients that appear in the above two equations are given by 
\begin{equation}
    \alpha_1=\frac{c_T^{2}}{2 H_i},\quad \alpha_2=\frac{\gamma-1}{\gamma^2 H_i^{2}}\left[c_{Si}^2-c_T^2\left(1-\gamma/2\right)\right],
    \label{eq12.1}
\end{equation}
\begin{equation}
    \alpha_3=\frac{c_{Sn}^2}{H_n}\left(1-\frac{H_n}{2 H_i}\right),\quad \alpha_4=\frac{c_{Sn}^2(\gamma-1)}{2\gamma H_nH_i},
    \label{eq12.3}
\end{equation}
\begin{equation}
    \alpha_5=\frac{c_{Sn}^2 c_T^{2}(\gamma -1)}{\gamma H_n v_A^{2}}\left(1-\frac{H_n}{H_i}\right), \quad  \alpha_6=\frac{c_{Sn}^2 \alpha_1}{\gamma^2 H_n v_A^{2}}(\gamma^2-3\gamma+2).
    \label{eq12.6}
\end{equation}

Equations (\ref{eq11})--(\ref{eq12}) describing the evolution of the two waves can be brought into a simpler form by introducing the reduced function for ions and neutrals of the form
\[
v_i(z,t)=Q_i(z,t)\exp(\lambda_i z), \quad  v_n(z,t)=Q_n(z,t)\exp(\lambda_n z).
\]
As the quantities $\lambda_i$ and $\lambda_n$ can be arbitrary, we can  choose their values so that first-order derivatives with respect to the spatial coordinate, $z$, vanish. Therefore, by choosing 
\begin{equation}
    \lambda_i=\frac{1}{4 H_i},\quad  \lambda_n=\frac{1}{4 H_i}-\frac{1}{2 H_n},
    \label{eq14}
\end{equation}
 the evolutionary equations (\ref{eq11})--(\ref{eq12}) can be represented as
\begin{equation} \label{eq16}
    \frac{\partial^2 Q_i}{\partial t^2}-c_T^2\frac{\partial^2 Q_i}{\partial z^2}+\omega_{i}^2 Q_i=0,
\end{equation}
\[
    \frac{\partial^2 Q_n}{\partial t^2}-c_{Sn}^{2} \frac{\partial^2 Q_n}{\partial z^2} +\Omega_{n}^2 Q_n+\nu_{ni} \frac{\partial Q_n}{\partial t}=
    \]
    \begin{equation}
    \label{eq17}
    \left(-\frac{c_{Sn}^2 c_T^{2}}{v_A^{2}}\frac{\partial^2 Q_i}{\partial z^2}+\delta Q_i \right)\exp\left(-\frac{z}{2\gamma h}\right),
\end{equation}
where now the coefficients $\omega_i^2$, $\Omega_n^2$ and $\delta$ are given by
\[
\omega_i^2=\left(\frac94-\frac{2}{\gamma}\right)\omega_{Ai}^2-\omega_{Ai}^2\frac{\beta\gamma}{2+\beta\gamma}\left(\frac32-\frac{2}{\gamma}\right)^2,
\]
\[
\Omega_n^2=\frac{c_{Sn}^2}{2c_{Si}^2}\omega_{Ai}^2+\omega_{An}^2+\frac{c_{Sn}^2}{4\gamma H_iH_n}(\gamma-2),
\]
and
\[
   \delta= \frac{c_{Sn}^{2}}{v_A^{2} (2+\gamma \beta)}\left[ \omega_{Ai}^{2} \left(\frac{1}{2}-\frac{2}{\gamma}+\frac{2}{\gamma^2}\right)+\frac{2c_{Si}^{2}}{c_{Sn}^2} \omega_{An}^{2}\left(1-\frac{2}{\gamma}+\frac{1}{\gamma^2}\right)+\right.
   \]
  \[ 
   \left. \frac{c_{Si}^{2}}{H_i H_n}\left(\frac14-\frac{3}{2 \gamma}+\frac{1}{\gamma^2}\right)\right],
\]
with $\beta=2c_{Si}^2/(\gamma \varv_{A}^2 )$ being the plasma-beta parameter, and  $\omega_{Ai}={c_{Si}}/{(2 H_i)}$ and $\omega_{An}={c_{Sn}}/{(2 H_n)}$ are the acoustics cut-off frequency for ions and neutrals.

The significance of the quantity $\omega_i$ in Equation (\ref{eq16}) becomes clear once a normal mode analysis is applied to this equation and consider that the function $Q_i(z,t)$ has a plane-wave dependence of the form $Q_i(z,t)\sim e^{i(kz-\omega t)}$. The resulting dispersion relation can be rearranged into the form
\begin{equation}
    k^2=\frac{\omega^2-\omega_i^2}{c_T^2}.
    \label{eq18}
\end{equation}
Propagating waves are possible only when the wavenumber, $k$, is real and this condition is satisfied if $k^2>0$, i.e. $\omega^2>\omega_i^2$. Therefore, waves will propagate only if their frequencies are larger than the cut-off value $\omega_i$, so the stratified solar atmosphere acts as a frequency filter, where only high frequency waves propagate. When $\omega<\omega_i$ waves will be evanescent with an e-folding length of $c_{Si}/\sqrt{\omega_i^2-\omega^2}$. The acoustic cut-off arises when ion-acoustic waves cannot propagate vertically because the wavelength is comparable with the density scale-height; consequently there is insufficient inertia on the low-density side of a compression to resist the acceleration of plasma, thereby cancelling too much of the pressure gradient to permit adequate subsequent compression of the surroundings, essential for causing the perturbation to propagate in a wave-like way. The dynamics operates on the vertical component of the motion, and is most effective for motion that is purely vertical. 

Equation (\ref{eq18}) can also be written as
\[
\frac{d^2 Q_i}{dz^2}+\frac{\omega^2-\omega_i^2}{c_{Si}^2}Q_i=0
\]
meaning that propagating/evanescent waves correspond to $d^2Q_i/dz^2<0$ and $d^2Q_i/dz^2>0$, respectively. 

A similar treatment for neutral-acoustic modes is not possible, and we will return to this aspect later. We should note here that in the strongly ionised limit the value of the ion cut-off frequency agrees (qualitatively) with the cut-off frequency for a fully ionised plasma, derived by, e.g. Rae and Roberts (1982).

The system of coupled equations (\ref{eq16})--(\ref{eq17}) describe the propagation of ion-acoustic and neutral-acoustic wave in space and time. All the coefficients that appear in homogeneous part of these equations are constants. The two partial differential equations will be solved as an initial value problem (IVP), where we aim to study the asymptotic evolution of waves.

\section{Asymptotic behaviour of guided slow waves}

In order to discuss the asymptotic behaviour of waves for large values of time we will need to solve the initial value problem associated with the two evolutionary equations (\ref{eq16})--(\ref{eq17}). To make analytical progress we will assume that all species have the same temperature, so that $T_e=T_i=T_n=T$. This assumption is in line with the physical requirement that a system will tend towards a state of equipartition of energy and uniform temperature that maximises the system's entropy. As a result, any local modification of temperature (and increase in the thermal speed of particles) is smoothed out after a few collisional times, i.e. over times that are smaller than the period of waves (very often this time is called the equilibration time) and any modifications in the distribution of particles is reduced in time, leading to a Maxwellian distribution. Since for the hydrogen plasma assumed here the mass of ions and neutrals are nearly identical, ions rapidly exchange energy with neutrals and tend to reach a thermal equilibrium with neutrals. Indeed, the amount of energy that is exchanged between ions and neutrals can be at most $m_im_n/(m_i+m_n)^2\approx 0.25$ times of their energy, making the process of thermalisation through collision very effective. In contrast, during the collision between electrons and hydrogen neutral atoms, electrons are able to transfer only $m_em_n/(m_e+m_n)^2\approx 5.4\times 10^{-4}$th part of their energy and it requires approximately 1850 collisions to reach the equipartition of energy between electrons and neutrals, and consequently, equality of their temperature. For a $T=10^4$ K plasma and a neutral number density of $n_n=2\times 10^{15}$ m$^{-3}$ the collisional frequency between electrons and neutrals is approximately 39 s$^{-1}$, meaning that in about 47 seconds the electron and neutral population reach a thermal equilibrium. 

Let us estimate the equilibration time between ions and neutrals. In the absence of flows and other spatial inhomogeneities, the evolution of the temperature is given by the energy equations written for the two species
\[
\frac{dT_i}{dt}=\nu_{in}(T_n-T_i),
\]
\[
\frac{dT_n}{dt}=\nu_{ni}(T_i-T_n).
\]
Assuming that the temperatures of the two species at the start of our investigation are ${\hat T}_i$ and ${\hat T}_i$, the temporal evolution of the temperatures with time (see Soler et al 2013) is given by
\[
T_i=T_f-({\hat T}_n-{\hat T}_i)\frac{\nu_{in}}{\nu_{in}+\nu_{ni}}e^{-(\nu_{in}+\nu_{ni})t},
\]
\[
T_n=T_f+({\hat T}_n-{\hat T}_i)\frac{\nu_{ni}}{\nu_{in}+\nu_{ni}}e^{-(\nu_{in}+\nu_{ni})t},
\]
where
\[
T_f=\frac{\nu_{in}{\hat T}_n+\nu_{ni}{\hat T}_i}{\nu_{in}+\nu_{ni}},
\]
is the final temperature the two species will tend to through collision. We can easily estimate the time ($t_f$) required for the two species to reach 99\% of the common temperature as
\[
t_f=\frac{1}{\nu_{ni}\left(1+\nu_{in}/\nu_{ni}\right)}\ln\left[10^2\frac{({\hat T}_i-{\hat T}_n)}{{\hat T}_i+{\hat T}_n\nu_{in}/\nu_{ni}}\right].
\]
Given the relationship between the two collisional frequencies we can write $t_f$ as
\[
t_f=\frac{1}{\nu_{ni}\left(1+n_n/n_i\right)}\left[\ln10^2\frac{({\hat T}_i-{\hat T}_n)}{{\hat T}_i+{\hat T}_n n_{n}/n_{i}}\right].
\]
Finally, taking into account that in the present study we deal with strongly ionised plasma for which $n_i\gg n_n$, the above relation simplifies to
\[
t_f\approx \frac{1}{\nu_{ni}}\left[4.6+\ln\left(1-\frac{{\hat T}_n}{{\hat T}_i}\right)\right].
\]
For an order of magnitude estimate let us consider that ${\hat T}_i=3{\hat T}_n$, and $\nu_{ni}=10$ s$^{-1}$. As a result, the time needed for the two species to reach 99\% of the thermal equilibrium is 0.4 seconds, i.e. thermal equilibrium between the massive particles is settled, indeed, very quickly. This conclusion is in line with the results obtained by earlier studies (e.g. Zaqarashvili et al. 2011, Soler et al. 2013 and Oliver et al. 2016).

As a consequence of the thermal equilibrium, the relationship between the sound speeds associated with the two constituent fluids becomes
\[
c_{Si}^2=\frac{\gamma (p_{0i}+p_{0e})}{\rho_{0i}}=\frac{\gamma k_B(T_{i}+T_e)}{m_i}=\frac{2\gamma k_BT_n}{m_n}=\frac{2\gamma p_{0n}}{\rho_{0n}}=2c_{Sn}^2.
\]
Using this result, the ratio of the propagation speed of waves associated to neutral and charged species takes the form
\begin{equation}
    \label{eq19}
    \frac{c_{Sn}^2}{c_T^2}=\frac{c_{Sn}^2}{c_{Si}^2}\left(1+\frac{c_{Si}^2}{v_A^2}\right)=\frac12\left(1+\frac{\gamma\beta}{2}\right)\approx \frac12,
\end{equation}
where we used the consideration that our investigation is valid for the low plasma-beta case. The above result shows that the wave associated to charged particles propagates with a speed that is roughly twice the propagation speed of neutral-acoustic mode. Another straightforward implication of the above assumption is that the gravitational scale-height of ions ($H_i$) is twice the scale height corresponding to neutrals ($H_n$), i.e. the density decrease of neutrals with height is faster than for ions. In addition, the reduced scale-height, $h$, defined by Equation  (\ref{eq10}), becomes $h=2H_n$. 

Because the two modes always appear together the above consideration raises an important aspect. Since the neutral acoustic modes are trailing the ion acoustic modes, the  former waves will propagate in an environment that is already modified by the ion acoustic mode and this materialises partly in a modified temperature and density that results from the perturbations caused by the ion acoustic modes. The passage of the ion acoustic mode will also modify the density of ions, and through collisions, the density of neutrals will also be modified. However, in the present study we will assume that these changes are insignificant and, therefore, will be neglected. It is likely that the correctness of our assumption can be checked only by rigorous numerical investigation. 

\subsection{Ion-acoustic modes}

Let us recall that the evolutionary equation for the charged fluid was obtained to be given by the Klein-Gordon equation 
\begin{equation} \label{eq30}
    \frac{\partial^2 Q_i}{\partial t^2}-c_T^2\frac{\partial^2 Q_i}{\partial z^2}+\omega_{i}^2 Q_i=0.
\end{equation}
We are going to consider the spatial positive domain and the solution of the above equation will be sought subject to the initial conditions $Q_i(z,0)=\partial Q_i(z,0)/\partial t=0$. In addition, we require that waves will vanish at $z\to \infty$, i.e. $Q_i(z\to \infty, t)=0$. 

The IVP problem can be studied by means of the Laplace transform. Accordingly, we introduce the Laplace transform of the function $Q_i(z,t)$ as
\begin{equation}
    \Psi_i(z,s)={\cal L}[Q_i(z,t]=\int_0^{\infty}Q_i(z,t)e^{-st}\;dt.
    \label{eq31}
\end{equation}
As a result, the Klein-Gordon equation for ions reduces
\begin{equation}
s^{2} \Psi_i(z,s)-c_T^{2} \frac{d^{2}}{d z^{2}} \Psi_i(z,s)+\omega_{i}^{2} \Psi_i(z,s)=0,
\label{eq32}
\end{equation}
that has to be solved subject to the boundary condition $\Psi_i(z\to \infty,s)=0$. The above equation can rearranged as
\begin{equation}
    \frac{d^{2}}{d z^{2}} \Psi_i(z,s)- \frac{s^{2}+\omega_{i}^{2}}{c_T^{2}} \Psi_i(z,s)=0,
    \label{eq33}
\end{equation}
whose general solution can be simply written as
\begin{equation}
    \Psi_i(z,s)= C_1 \exp \left(\frac{z}{c_T}\sqrt{s^{2}+\omega_{i}^{2}}\right) + C_2\exp \left(-\frac{z}{c_T}\sqrt{s^{2}+\omega_{i}^{2}}\right),
    \label{eq34}
\end{equation}
where $C_1$ and $C_2$ are arbitrary constants. Clearly, the first term will not satisfy the required boundary condition, therefore we choose $C_1=0$. Let us consider that at $z=0$ the wave is driven by a function $Q_i(0,t)={\cal A}_0(t)$ and its Laplace transform is $\Psi_i(0,s)=a_0(s)$. After applying this condition to the  general solution, we obtain
\begin{equation}
     \Psi_i(z,s)= a_0(s)\exp \left(-\sqrt{\frac{s^{2}+\omega_{i}^{2}}{c_T^{2}}}z \right).
     \label{eq35}
\end{equation} 
Now, the function $Q_i(z,t)$ can be obtained by applying the inverse Laplace transform to the function given by Equation (\ref{eq35}). Since we have to compute the inverse Laplace transform of a product, we will use the convolution theorem. In finding the value of the inverse Laplace transform we will closely follow the method outlined by Sutmann et al. (1998). 

In order to find the inverse Laplace transform of Equation (\ref{eq35}) we use the identity (Bateman and Erd\'elyi 1954)
\begin{equation}
\label{eq36}
{\cal L}^{-1}\left[\frac{e^{-a\sqrt{s^2+\omega_i^2}}}{\sqrt{s^2+\omega_i^2}}\right]=\left\{\begin{array}{ll}
J_0\left(\omega_i\sqrt{t^2-a^2}\right), & \text{for }  t> a\\
0, & \text{for } 0< t< a,\\
\end{array}\right.
\end{equation}
where $J_0$ is the zero-th order Bessel function. Let us define the function
\begin{equation}
    \label{eq37}
    I=\frac{e^{-a\sqrt{s^2+\omega_i^2}}}{\sqrt{s^2+\omega_i^2}}=\int_a^{\infty}J_0\left(\omega_i\sqrt{t^2-a^2}\right)e^{-st}dt.
\end{equation}
We differentiate both sides of Equation (\ref{eq37}) with respect $a$, so that
\[
    \frac{d I}{d a}= 
    \]
    \begin{equation}
    -\omega_i a \int_{a}^{\infty} \frac{J_0^{'}\left(\omega_i \sqrt{t^{2}-a^{2}}\right)}{\sqrt{t^{2}-a^{2}}} e^{-st}dt - e^{-as}=- \exp\left(-a\sqrt{\omega_i^{2}+s^{2}}\right).
    \label{eq38}
\end{equation}
We can use the identity $J_0^{'}(x)=-J_1(x)$, and  substitute $a$ by $t_i=z/c_T$ to obtain
\begin{equation}\label{eq39}
    \exp\left(-t_i\sqrt{\omega_{i}^{2}+s^{2}}\right )=\exp(-s t_i)-\omega_{i}t_i \int_{t_i}^{\infty} \frac{J_1 \left(\omega_{i} \sqrt{t^{2}-t_i^{2}}\right)}{\sqrt{t^{2}-t_i^{2}}} e^{-s t} dt.
\end{equation}
Now, introducing Equation (\ref{eq39}) into Equation (\ref{eq35}), we find that
\begin{equation}\label{eq40}
    \Psi_i(z,s)= a_0(s) \exp(-s t_i)- a_0(s)\omega_{i}t_i \int_{t_i}^{\infty} \frac{J_1 \left(\omega_i \sqrt{t^{2}-t_i^{2}}\right)}{\sqrt{t^{2}-t_i^{2}}} e^{-s t} dt .
\end{equation}
Note that the z-dependence of the above function is ensured through the expression of $t_i$, which was introduced to simplify the notation. Let us define the function
\begin{equation}
    Z_i(z,t)=-\omega_{i}t_i \frac{J_1 \left(\omega_{i} \sqrt{t^{2}-t_i^{2}}\right)}{\sqrt{t^{2}-t_i^{2}}} H\left(t-t_i\right),
    \label{eq41}
\end{equation}
where the $H(t-t_i)$ is the Heaviside step function.

After applying the second shifting theorem to the first term on the right-hand side of Equation (\ref{eq40}), we obtain
\begin{equation}
   a_0(s) \exp(-s t_i) = {\cal L}\left[{\cal A}_0\left(t-t_i\right) H\left(t-t_i\right)\right].
   \label{eq42}
\end{equation}
As a result, Equation (\ref{eq40}) becomes
\begin{equation}\label{eq43}
    \Psi_i(z,s)={\cal L}\left[{\cal A}_0 (t-t_i) H(t-t_i)\right]+ {\cal L}\left[{\cal A}_0(t) Z_i(z,t)\right].
\end{equation}
Using the convolution theorem, the second term in right-hand side of above equation can be written as
\begin{equation}
    {\cal L}\left[{\cal A}_0(t) Z_i(z,t)\right]= {\cal L}\left[\int_{0}^{t} {\cal A}_0(t-\tau) Z_i(z,\tau) d \tau\right].
    \label{eq44}
\end{equation}

\noindent Since the original function $Q_i(z,t)$ can be determined as the inverse Laplace transform of the function $\Psi_i(z,s)$ given by Equation (\ref{eq43}), eventually we we obtain 
\begin{equation}
Q_i(z,t)={\cal A}_0 \left(t-t_i\right) H\left(t-t_i\right)+\int_{0}^{t} {\cal A}_0(t-\tau) Z_i(z,\tau) d\;\tau.
\label{eq45}
\end{equation}
In the case of ion-acoustic modes the spatial and temporal evolution of the reduced speed, $Q_i(z,t)$ is given by Equation (\ref{eq45}). Given the specific driver we have 
\[
{\cal A}_0(t-t_i)=V_0[H(t-t_i)-H(t-t_i-P)]e^{i\omega(t-t_i)}.
\]
Since we are interested in the asymptotic behaviour of waves it is clear that $t\gg t_i$, which implies $t\gg (t_i+P)$. As a result both Heaviside functions become unity, and the first term of Equation (\ref{eq45}) becomes zero. Further, the second term of Equation (\ref{eq45}) can be written as
\[
Q_i(z,t)=V_0\int_0^tH(t-\tau)e^{-i\omega(t-\tau)}Z_i(z,\tau)d\tau-
\]
\[
-V_0\int_0^tH(t-\tau-P)e^{-i\omega(t-\tau)}Z_i(z,\tau)d\tau.
\]
It is clear that the first term cancels because all the values of $\tau$ have to be in the interval $(0,t)$, for which the Heaviside function is zero. Using the Heaviside function, the reduced speed, $Q_i$ can be written as
\[
Q_i(z,t)=-V_0\int_{t-P}^t e^{-i\omega(t-\tau)}Z_i(z,\tau)d\tau.
\]
In order to make analytical progress we will rewrite the convolutive integral such that 
\[
\int_{t-P}^{t}\dots d\tau=\int_{t-P}^{\infty}\dots d\tau-\int_t^{\infty}\dots d\tau.
\]
In order to estimate the value of these integrals we should keep in mind that the asymptotic analysis is valid provided $t\gg t_i$ or $\tau\gg t_i$ for which the Bessel function $J_1(x)$ for large arguments can be written as
\[
J_1(x)\approx \frac{2}{\sqrt{\pi x}}\left[\cos\left(x-\frac{3\pi}{4}\right)+{\cal O}\left(\frac{1}{x}\right)\right].
\]
After some straightforward calculations (see Sutmann et al. 1998, Appendix B) we eventually obtain
\[
    Q_i(z,t)=V_0\sqrt{\frac{2\omega_i}{\pi}}\frac{1}{\omega^2-\omega_i^2}\frac{2t_i}{t^{3/2}}\sin\left(\frac{\omega_iP}{2}\right)\times
    \]
    \begin{equation}
     \label{eq60}
        \left[\omega_i\sin\left(\omega_i (t-P/2)-\frac{3\pi}{4}\right)-i\omega\cos\left(\omega_i (t-P/2)-\frac{3\pi}{4}\right)\right].
    \end{equation}
Clearly this solution describes a wave whose transient part that oscillates with the cut-off frequency, $\omega_i$, but this decays in time as $t^{-3/2}$. As a result, an observer situated at a given height, $z_0$, would observe a damped slow wave propagating with the cut-off frequency $\omega_i$ and free oscillations (the steady solution) are not present.


\subsection{Neutral-acoustic modes} 

The equation that describes the spatial-temporal evolution of these waves is given by Equation (\ref{eq17}). It is clear that the evolution of these waves (described by the left-hand side of Equation \ref{eq17}) is driven by ions. In contrast to ions, where in the first order of approximation the collisions with neutrals can be neglected, in the case of neutrals the collisions with ions will play an essential role, and this effect is described by the last term on the left-hand side of Equation (\ref{eq17}). This equation is an inhomogeneous partial differential equation and solutions can be obtained by determining the complementary solution and a particular solution that is driven by the form of the inhomogeneous term. The complementary solution can be obtained after solving the equation  \begin{equation}\label{eq45.1}
    \frac{\partial^2 Q_n}{\partial t^2}-c_{Sn}^{2} \frac{\partial^2 Q_n}{\partial z^2} +\Omega_{n}^2 Q_n+\nu_{ni} \frac{\partial Q_n}{\partial t}=0.
    \end{equation}
The above equation is the well-known telegrapher's equation that can be easily reduced to a Klein-Gordon equation. Accordingly, let us introduce a new function so that $Q_n (z,t)=q_n (z,t)e^{-\nu_{ni} t/2}$ As a result the equation that describes the complementary solution of neutral-acoustic modes becomes
\begin{equation}
\label{eq46}
    \frac{\partial^2 q_n}{\partial t^2}-c_{Sn}^{2} \frac{\partial^2 q_n}{\partial z^2} +\left(\Omega_{n}^2-\frac{\nu_{ni}^2}{4}\right) q_n=0.
\end{equation}
It can be shown that the quantity $\Omega_{n}^2-\nu_{ni}^2/4$ is always negative. Again, using a normal mode analysis similar to the method employed in the case of ion-acoustic modes, it becomes clear that neutral-acoustic modes propagate with no cut-off. 

Now, let us write the governing equation for the neutral-acoustic mode in the form
\[
    \frac{\partial^2 q_n}{\partial t^2}-c_{Sn}^{2} \frac{\partial^2 q_n}{\partial z^2} -\omega_n^2 q_n=
    \]
    \begin{equation}
    \label{eq47}
   = \left(-\frac{c_{Sn}^2c_T^2}{v_A^2}\frac{\partial^2 Q_i}{\partial z^2}+\nu_{ni}\frac{\partial Q_i}{\partial t}+\delta Q_i\right)e^{\nu_{ni}t/2}e^{-z/4\gamma H_n},
    \end{equation}
where $\omega_n^2=\nu_{ni}^2/4-\Omega_n^2$. Next, we apply the Laplace transform to the above equation and denote the Laplace transform of the function $q_n$ as
\[
\Psi_n(z,s)=\int q_n(z,t)e^{-st}\;dt.
\]
Using the expression of $\Psi_i(z,s)$ given by Equation (\ref{eq35}) we can write the governing equation for neutrals as
\begin{equation}
\frac{\partial^2 \Psi_n}{\partial z^2}-\frac{s^2-\omega_n^2}{c_{Sn}^2}\Psi_n=f(z,s),
\label{eq48}
\end{equation}
where, with the help of the shifting theorem, the inhomogeneous part, $f(z,s)$ is given by
\[
    f(z,s)=\left\{-\frac{1}{v_A^2}\left[\left(s-\frac{\nu_{ni}}{2}\right)^2+\omega_i^2\right]+\frac{\nu_{ni}}{c_{Sn}^2}\left(s-\frac{\nu_{ni}}{2}\right)+\frac{\delta}{c_{Sn}^2}\right\}\times
    \]
    \begin{equation}
     \label{eq49}
    a_0\left(s-\frac{\nu_{ni}}{2}\right)exp\left[-t_i\sqrt{\left(s-\frac{\nu_{ni}}{2}\right)^2+\omega_i^2}-\frac{z}{4\gamma H_n}\right].
\end{equation}
The solution of the homogeneous part of the Equation (\ref{eq48}) that satisfies the condition at infinity becomes
\begin{equation}
    \label{eq50}
    \Psi_n^{hom}=B_1\exp\left[-\frac{z}{c_{Sn}}\sqrt{s^2-\omega_n^2}\right],
\end{equation}
and the value of the constant $B_1$ will be chosen such that its value will be the Laplace transform of the driver at $z=0$. For simplicity we will assume that the waves associated to both fluids are initiated by the same driver, therefore we will write $B_1=a_0(s)$. 

To find the inverse Laplace transform of the homogeneous solution we use the identity (Bateman and Erd\'elyi 1954)
\[
{\cal L}^{-1}\left[\frac{e^{-a\sqrt{s^2-\omega_n^2}}}{\sqrt{s^2-\omega_n^2}}\right]=\left\{\begin{array}{ll}
I_0\left(\omega_n\sqrt{t^2-a^2}\right), & \text{for }  t> a\\
0, & \text{for } 0< t< a,\\
\end{array}\right.
\]
where $I_0(x)$ is the modified Bessel function of order zero. Now let us define the function
\begin{equation}
    \label{eq51}
    J=\frac{e^{-a\sqrt{s^2-\omega_n^2}}}{\sqrt{s^2-\omega_n^2}}=\int_a^{\infty}I_0(\omega_n\sqrt{t^2-a^2})e^{-st}\;dt.
\end{equation}
After differentiating the above function with respect to $a$, we obtain
\[
\frac{dJ}{da}=-e^{-a\sqrt{s^2-\omega_n^2}}=-a\omega_n\int_a^{\infty}\frac{I_0^{\prime}(\omega_n\sqrt{t^2-a^2})}{\sqrt{t^2-a^2}}e^{-st}\;dt-e^{-as},
\]
where {\it dash} denotes the derivative of the function $I_0(x)$ with respect to its argument. Using the identity $I_0^{\prime}(x)=I_1(x)$ and replacing $a$ by $t_n=z/c_{Sn}$ we obtain
\begin{equation}
    \label{eq52}
    \exp\left(-t_n\sqrt{s^2-\omega_n^2}\right)=\omega_n t_n\int_{t_n}^{\infty}\frac{I_1(\omega_n\sqrt{t^2-t_n^2})}{\sqrt{t^2-t_n^2}}e^{-st}dt+e^{-st_n}.
\end{equation}
It can be easily shown that in the low beta approximation $t_i\approx t_n/\sqrt{2}$. Let us define the function 
\[
Z_n(z,t)=\omega_n t_n\frac{I_1(\omega_n\sqrt{t^2-t_n^2})}{\sqrt{t^2-t_n^2}}H\left(t-t_n\right).
\]
As a result, the solution of the homogeneous part of the governing equation for neutral-acoustic slow waves becomes
\[
q_n(z,s)=a_0(s)e^{-st_n}+a_0(s){\cal L}\left[Z_n(z,t)\right].
\]
After applying the inverse Laplace transform and the convolution theorem, the solution becomes 
\begin{equation}
    \label{eq53}
    q_n(z,t)={\cal A}_0\left(t-t_n\right)H\left(t-t_n\right)+\int_0^t\;{\cal A}_0(t-\tau)Z_n(z,\tau)\;d\tau.
\end{equation} 

In order to determine the particular solution of the evolutionary equation for neutrals, we will need to calculate the inverse Laplace transform of the expression
\begin{equation}
    \label{eq54}
    D(z,s)=a_0\left(s-\frac{\nu_{ni}}{2}\right)K(s)e^{-z/4\gamma H_n}\exp\left[-t_i\sqrt{(s-\nu_{ni}/2)^2+\omega_i^2}\right],
\end{equation}
where the function $K(s)$ is defined as
\[
K(s)=\frac{-\frac{\beta\gamma}{4}\left[\left(s-\frac{\nu_{ni}}{2}\right)^2+\omega_i^2\right]+\nu_{ni}\left(s-\frac{\nu_{ni}}{2}\right)+\delta}{s^2-\omega_n^2-c_{Sn}^2\left[\frac{1}{4\gamma H_n}+\frac{1}{c_T}\sqrt{\left(s-\frac{\nu_{ni}}{2}\right)^2+\omega_i^2}\right]^2}
\]
The above relation shows that we will need to deal with the inverse Laplace transform of a triple product, therefore we will use the triple convolution formula. According to the standard definition if $F(s)$, $G(s)$ and $H(s)$ are the Laplace transforms of the functions $f(t)$, $g(t)$ and $h(t)$, then
\[
{\cal L}^{-1}\left[F(s)G(s)H(s)\right]=\int_0^\tau\left[f(t-\tau)\int_0^{\tau}g(\tau-\zeta)h(\zeta)d\zeta\right]d\tau.
\]
Since the inverse Laplace transform of the exponential term in Equation (\ref{eq54}) has already been obtained (see Equation \ref{eq44}), the only task here will be to derive the inverse Laplace transform of the function $K(s)$. This function has two simple poles at the zeros of the denominator, therefore the inverse Laplace transform can be obtained as the sum of the residues at the two poles. It is easy to see that the denominator is singular at
\begin{equation}
    \label{eq55}
    \Gamma_{1,2}=\frac{-\nu_{ni}\pm {\cal G}}{1-g/2\omega_ic_T},
\end{equation}
where
\[
{\cal G}=\left[\nu_{ni}^2-\left(2-\frac{g}{\omega_ic_T}\right)\left(\Omega_n^2-\frac{g^2}{16c_{Sn}^2}-\frac{\omega_i^2}{2}-\frac{\omega_ig}{2c_T}\right)\right]^{1/2}.
\]
It can be shown that for typical chromospheric conditions ${\cal G}$ is real, therefore both roots, $\Gamma_{1,2}$, are real and negative. As a result, the inverse Laplace transform of the function $K(s)$ becomes
\begin{equation}
    \label{eq56}
    {\cal L}^{-1}[K(s)]=\frac{i\pi\left(1-g/2\omega_ic_T\right)}{{\cal G}}e^{\nu_{ni}t/2} \left(y_1e^{\Gamma_1t}-y_2e^{\Gamma_2t}\right).
    \end{equation}
with 
\[
y_j=\delta-\frac{\beta\gamma}{4}\left(\omega_i^2+\Gamma_j^2\right)+\nu_{ni}\Gamma_j, \quad j=1,2.
\]
Taking into account the inverse Laplace transform of all terms that appear in the expression of $D(z,s)$ given by Equation (\ref{eq54}) after a lengthy, but straightforward calculation we can obtain the the particular solution of Equation (\ref{eq47}). However, since the expression of the whole particular solution is far too long and the expression of this solution will not be used in the present form, we choose to give the detailed solution once the asymptotic expression for large values of time is derived.

The asymptotic solution of these equations refer to the case of large values of time, i.e. for values of time for which $t\gg z/c_T$. Given the relationship between the propagation speed of the two modes, this condition includes the condition we impose for neutral-acoustic modes.

\subsection{Oscillations driven by a sinusoidal pulse}

We choose to drive the system (both species) with a harmonic pulse of the form ${\cal A}_0(t)=V_0e^{-i\omega t}[H(t)-H(t-P)]$, where $P=2\pi/\omega$. This driver acts for a duration $P$, after which is stopped. The driver acts at $z=0$. In what follows we are going to discuss separately the asymptotic solution for both species. 

Now let us return to neutral acoustic modes, whose evolutionary equation is given by Equation (\ref{eq47}). First, let us investigate the asymptotic form of the homogeneous solution given by Equation (\ref{eq53}). Again, assuming the same harmonic driver of the form ${\cal A}_0(t)=V_0e^{-i\omega t}[H(t)-H(t-P)]$ situated at $z=0$ we have
\[
    \label{eq61}
    q_n(z,t)=V_0e^{-i\omega\left(t-t_n\right)}[H\left(t-t_n\right)-H(t-P-t_n)]+
    \]
    \[
   +V_0\int_0^t H(t-\tau)e^{-i\omega(t-\tau)}Z_n(z,\tau)\;d\tau-
   \]
   \begin{equation}
   -V_0\int_0^t H(t-\tau-P)e^{-i\omega(t-\tau)}Z_n(z,\tau)\;d\tau.
\end{equation}

Similar to the discussion presented in the case of ion-acoustic slow modes the contributions of the first two terms of the above equation are zero. As a result after taking into account the restriction imposed by the Heaviside function, the homogeneous part of the equation of $q_n$ is given by 
\[
q_n(z,t)=-V_0\int_{t-P}^t e^{-i\omega(t-\tau)}Z_n(z,\tau)\;d\tau.
\]
Since we are investigating the asymptotic behaviour of waves for large values of time, we can write that this corresponds to $\tau\gg t_n$, which means that our equation reduces to
\begin{equation}
    \label{eq62}
    q_n(z,t)=-V_0e^{-i\omega t}\int_{t-P}^t\frac{I_1\left(\omega_n\tau\right)}{\tau}e^{i\omega\tau}\;d\tau.
\end{equation}
For large arguments the modified Bessel function can be written as
\[
I_1(\omega_n\tau)\approx \frac{e^{\omega_n\tau}}{(2\pi\omega_n\tau)^{1/2}}.
\]
Therefore the evolutionary equation for the homogeneous part of the governing equation for neutrals becomes
\begin{equation}
    \label{eq63}
   q_n(z,t)=-\frac{V_0e^{-i\omega t}}{\sqrt{2\pi\omega_n}}\int_{t-P}^t\frac{e^{(\omega_n+i\omega)\tau}}{\tau^{3/2}}\;d\tau.
\end{equation} 
The integral in the above relation can be given approximately (see Appendix A). As a result the evolution of the homogeneous part of $q_n(z,t)$ becomes
\begin{equation}
    \label{eq64}
    q_n(z,t)=-\frac{V_0(\omega_n-i\omega)e^{\omega_nt}}{\sqrt{2\pi\omega_n}t^{3/2}(\omega_n^2+\omega^2)}\left[1-e^{(\omega_n+i\omega)P}\right],
\end{equation}
where we used the approximation
\begin{equation}
\frac{1}{(t-P)^{3/2}}\approx \left(1+\frac32\frac{P}{t}\right)\frac{1}{t^{3/2}}=\frac{1}{t^{3/2}}+{\cal O}\left(t^{-5/2}\right).
\label{eq64.1}
\end{equation}
Now taking into account the relationship between $q_n(z,t)$ and $Q_n(z,t)$ we can find that the homogeneous solution of the evolutionary equation for neutrals becomes
\begin{equation}
    \label{eq65}
    Q_n^{hom}=-V_0\sqrt{\frac{\omega_n}{2\pi}}\frac{\omega_n-i\omega}{\omega_n^2+\omega^2}\frac{e^{(\omega_n-\nu_{ni}/2)t}}{t^{3/2}}\left[1-e^{-(\omega_n+i\omega)P}\right].
\end{equation}
Since $\nu_{ni}/2>\omega_n$, it is clear that the above solution describes an evanescent wave whose amplitude decays very rapidly due to collisions.

Finally, using the technique presented earlier, the inverse Laplace transform of the inhomogeneous part of Equation (\ref{eq54}) that gives the particular solution of Equation (\ref{eq47}) is
\[
Q_n^{inh}(z,t)=\frac{A_2A_3V_0}{t^{3/2}}\times
\]
\[
\left[(\omega\sin\Phi_1-i\omega_i\cos\Phi_1)\left(\frac{y_1\Gamma_1}{\Gamma_1^2+\omega_i^2}-\frac{y_2\Gamma_2}{\Gamma_2^2+\omega_i^2}\right)-\right.
\]
\[
\left.-(\omega\sin\Phi_2-i\omega_i\cos\Phi_2)\left(\frac{y_1\Gamma_1e^{\Gamma_1P}}{\Gamma_1^2+\omega_i^2}-\frac{y_2\Gamma_2e^{\Gamma_2 P}}{\Gamma_2^2+\omega_i^2}\right)-\right.
\]
\[
\left.-\omega_i(\omega\cos\Phi_1-i\omega_i\sin\Phi_1)\left(\frac{y_1}{\Gamma_1^2+\omega_i^2}-\frac{y_2}{\Gamma_2^2+\omega_i^2}\right)+\right.
\]
\begin{equation}
\label{eq66}
\left.+\omega_i\left(\omega\cos\Phi_2-i\omega_i\sin\Phi_2\right)\left(\frac{y_1e^{\Gamma_1P}}{\Gamma_1^2+\omega_i^2}-\frac{y_2e^{\Gamma_2P}}{\Gamma_2^2+\omega_i^2}\right)\right]
\end{equation}
where we used the notations
\[
A_2=\frac{\pi e^{-z/4\gamma H_n}\left(1-g/2\omega_i c_T\right)\sin \left(\omega_iP/2\right)}{{\cal G}},
\]
\[
A_3=\sqrt{\frac{2\omega_i}{\pi}}\frac{2}{\omega^2-\omega_i^2}t_i, 
\]
\[
\Phi_1=\omega_i\left(t-\frac{P}{2}\right)-\frac{3\pi}{4},\quad \Phi_i=\omega_i\left(t-\frac{3P}{2}\right)-\frac{3\pi}{4}.
\]
In contrast to the homogeneous solution, the particular solution shows a decaying oscillatory motion with the cut-off frequency of ions. This behaviour is a consequence of the coupling between neutrals and ions, where ions provide the oscillatory background for neutrals and the oscillatory behaviour of neutrals is driven by ions via collisions. 

\section{Application to solar atmosphere}

In what follows we are going to analyse our results assuming typical solar chromsopheric values for density and temperature. For magnetic field we assume a field strength of 10 G throughout all our investigations. 

In order to estimate key parameters that are important for our calculations we are going to consider that the plasma has a temperature of $T=10^4$ K and the number densities of ions and neutrals are $n_i=2\times 10^{15}$ m$^{-3}$ and $n_n=2\times 10^{13}$ m$^{-3}$. With these parameters we can estimate that the characteristic speeds will be $c_{Si}=16.6$ km s$^{-1}$, $c_{Sn}=11.7$ km s$^{-1}$ and $v_A=450$ km s$^{-1}$, which would result in a plasma- $\beta=1.7\times 10^{-3}$ and a cusp speed on the charged fluid $c_T=15.58$ km s$^{-1}$. For the given density and temperature values the collisional frequency between neutrals and ions can be determined with the help of Equations (\ref{eq6.1})--(\ref{eq6.2}) and results in  $\nu_{ni}=10.48$ s$^{-1}$. Finally, the gravitational scale-heights connected to ions and neutrals in thermal equilibrium become $H_i=2H_n=0.5$ Mm

Our analysis showed that ion-acoustic modes propagate in the stratified plasma such that their frequency is affected by a cut-off value. Using the definition of this quantity given by Eq. (\ref{eq16}) we obtain that for the representative temperature we have chosen, $\omega_i\approx 0.015$ Hz and it varies as $T^{-1/2}$ ( we should mention here that the value of the cut-off we would obtain for a fully ionised plasma for the same values of temperature and magnetic field, would be almost identical with the above value thanks to the strongly ionised limit employed by us). In addition, the variation of the cut-off frequency with respect to plasma-beta shows a very weak dependence. It is interesting to note that Leake et al. (2005) found that Alfv\'en waves have a cut-off  frequency of 0.6 Hz. As we proved earlier in Section 4, neutral-acoustic modes propagate with no cut-off frequency, however employing a normal mode analysis (i.e. assume that perturbations are proportional to the exponential factor $e^{i(kz-\omega t)}$  the homogeneous part of Eq. (\ref{eq47}) reduces to $\omega^2=k^2c_{Sn}^2-\omega_n^2$, so the requirement of propagating wave ($\omega^2>0$) means that in the case of neutral-acoustic waves the condition $k>\omega_n/c_{Sn}\approx \nu_{ni}/2c_{Sn}$ has to be satisfied (here the stratification effects are much smaller). Since $\omega_n$ depends on collisional frequency, the wave-number cut-off will be influenced by collisions. For the values of characteristic speeds and collisional frequency determined earlier neutral-acoustic waves will propagate provided their wavenumber is larger than $5\times 10^{-4}$ m$^{-1}$, or their wavelength is shorter than $1.25\times 10^4$ m. Clearly, such small wavelengths are impossible to observe with the current observational facilities. That is why observations can detect only one mode (connected to the charged species), while the neutral-acoustic modes remain sub-resolution modes. Such condition connected to wavenumbers is not imposed on ion-acoustic modes, for these waves the only restriction remains that their frequency has to be larger than the cut-off frequency $\omega_i$. 

If the above conditions are not satisfied, neutral-acoustic modes are becoming non-propagating entropy modes, i.e. modes whose frequency is purely imaginary. In the case of these modes all perturbations are zero, except density and temperature perturbations in such a way that the pressure perturbation is constant. Entropy mode own their existence to the collisions of neutrals with ions in strongly ionised limit and they play important role in the development and evolution of turbulences in the presence of small spatial scales (see, e.g. Lithwick and Goldreich, Soler et al. 2013).

Now let us return to the study of the temporal evolution of the two waves. For that we are going to fix the value of height and study the temporal evolution of the reduced velocity for the two waves. 

\begin{figure}
   \includegraphics[width=\columnwidth]{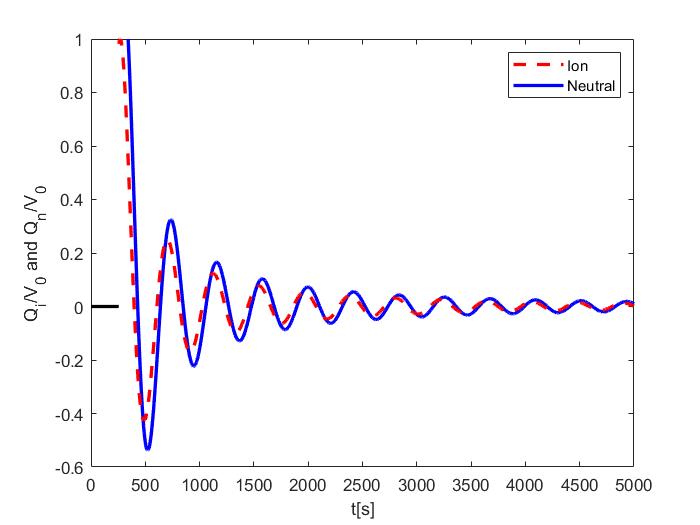}
      \caption{The temporal evolution of neutral-acoustic (solid lines) and ion-acoustic (dashed lines) modes at $z=4$ Mm. The slow sausage modes associated with the two species is driven by a sinusoidal pulse of lifetime $P$. Both slow modes oscillate with the ion cut-off frequency, $\omega_i$. For an observer situated at the observational height of 4 Mm, wave-like behaviour will be observable only after the delay time $t_i=z/c_T$. The delay time is shown here as a horizontal straight line.  }
   \end{figure}
In Fig. 2 we plot the temporal evolution of neutral-acoustic (solid line) and ion-acoustic (dashed line) slow mode at a given height ($z=4$ Mm) as given by the real parts of Eqs. (\ref{eq60}) and (\ref{eq66}). Due to the coupling between the two species, both waves oscillate with the same frequency $\omega_i$. It clear that the neutral-acoustic mode has a larger amplitude and decays slower than the corresponding ion-acoustic modes.  The two modes are excited at the $z=0$ level with the driving frequency $\omega=0.1$ Hz. Since the lifetime of the driver is limited (here chosen to be $P=20\pi$ s), the free oscillations associated with the two species are absent, instead of both slow modes attenuate. However, we should keep in mind that this attenuation is not due to physical damping (here collisions), instead it is due to dispersion and expansion of the cross section of the magnetic flux tube.

\section{Conclusions}

   Our study was devoted to the investigation of the temporal and spatial evolution of slow sausage waves propagating in an expanding magnetic flux tube in a gravitationally stratified atmosphere. The plasma temperatures are typical for the solar chromosphere, where the ionisation degree of the plasma is high, nevertheless the plasma is not fully ionised. Given the very different concentration of neutrals and charged species, the ratio between neutral and charged density is very small and this ratio was used as a small parameter in deriving the evolutionary equation for waves. The plasma was assumed to be isothermal,  which implies that all characteristic speeds are constant quantities.
   
   The evolutionary equation for slow sausage waves associated with the two species was derived in the linear limit. While the equation for waves associated with the charged particles is described by a Klein-Gordon equation, for neutrals this becomes the telegrapher's equation. Given the plasmas's high degree of ionisation the collisions have different role for the two species. For ions the collision with neutrals is just a secondary effect (and proportional to the density ratio between neutrals and ions). As a result the equation for ions (in the leading order) is not affected by collisions. In contrast, the equation for the neutrals species is strongly affected by the collisions between neutrals and ions, causing a strong decay of waves. While the ion-related waves propagate with a cut-off frequency, neutral sausage modes propagate with no frequency cut-off thanks to the collisions between species. In contrast, propagating slow waves associated to neutrals are possible only for wavelengths that are shorter than 12.5 km, that is they are small wavelength waves. 
   
   The evolutionary equations have been solved as an initial value problem, imposing a oscillatory pulse driver and an atmosphere that is unbounded in the $z$ direction. We considered the situation when the pulse has already passed through the atmosphere (i.e. we performed an asymptotic analysis valid for $t>>z/cT$), which implies that an observer would just observe the wake left behind the pulse. This wake oscillates with the cut-off frequency of the ion population. In other words, steady oscillations are excluded, and the system will oscillate with the transient part of the solution that decays as $t^{-3/2}$. This result is similar to the findings of Kalkofen et al. (1994) and Sutmann et al. (1998). 
   
   Slow sausage waves associated with neutrals propagate with no cut-off but given the high degree of coupling with ions, these will impose on neutrals the same behaviour, i.e. the transient solution of neutral slow wave oscillate with the same ion-related cut-off frequency and show the same temporal damping pattern as in the case of ions.  It is very likely that in strongly ionised plasmas these waves will have a very rapid decay, even in the absence of the simplifications we imposed to the employed model. That would mean that any possible observation of these waves has to be carried out in an environment where the ionisation degree is moderate. The presence of the cut-off frequency for ion-acoustic waves also implies that for a driving frequency smaller than the cut-off frequency, the ion-acoustic mode becomes evanescent (exponentially decaying), while the slow waves associated with neutrals will still propagate unaffected. This has large ranging consequences for observation of waves in the solar atmosphere.  Finally we should mention that when oscillations are driven by a sinusoidal pulse, whose frequency is identical with the ion cut-off frequency, the slow sausage modes associated to the two species will not propagate as these are free oscillations (for details see, e.g. Sutmann et al. 1998).
   
   Any attempt to describe wave propagation in a different plasma and field environments would require a detailed numerical analysis of the coupled system of charged particles and neutrals
   
   \section*{Acknowledgements}
      AA acknowledges Umm-AlQura University and Ministry of Education in the Kingdom of Saudi Arabia for their financial support. IB, VF and GV is grateful to The Royal Society, International Exchanges Scheme, collaboration with Brazil (IES191114) and Chile (IE170301). VF would like to thank the International Space Science Institute (ISSI) in Bern, Switzerland, for the hospitality provided to the members of the team on `The Nature and Physics of Vortex Flows in Solar Plasmas'. This research has received partial financial support from the European Union’s Horizon 2020 research and innovation program under grant agreement No. 824135 (SOLARNET). The authors are grateful for the anonymous Referee for his/her comments and suggestions.






\section*{Data Availability}

The data underlying this article will be shared on reasonable request to the corresponding author.




\bibliographystyle{mnras}




\begin{appendix}

\section{Evaluation of the integral in Equation (52)}

The value of the integral that is given in Equation (\ref{eq63}) can be given in approximate form for large values of $\tau$. The integral we have to estimate is 
\begin{equation}
    \label{A2}
    R(z,t)=\int_{t-P}^t\frac{e^{(\omega_n+i\omega)t}}{\tau^{3/2}}\;d\tau.
\end{equation}
Using integration by parts we have
\[
R(z,t)=\left.\frac{1}{\omega_n+i\omega}\frac{e^{(\omega_n+i\omega)\tau}}{\tau^{3/2}}\right|_{t-P}^t+\frac{3}{2(\omega_n+i\omega}\int_{t-P}^t\frac{e^{(\omega_n+i\omega)t}}{\tau^{5/2}}\;d\tau=
\]
\[
=\frac{1}{\omega_n+i\omega}\frac{e^{(\omega_n+i\omega)\tau}}{\tau^{3/2}}\left[1+\frac{3}{2(\omega_n+i\omega)\tau}\right]_{t-P}^t+
\]
\[
\frac{15}{4(\omega_n+i\omega)^2}\int_{t-P}^t\frac{e^{(\omega_n+i\omega)t}}{\tau^{7/2}}\;d\tau.
\]
The above relation can be re-arranged into
\[
\int_{t-P}^t\frac{e^{(\omega_n+i\omega)t}}{\tau^{3/2}}\left(1-\frac{15}{2\tau(\omega_n+i\omega)}\right)\;d\tau=
\]
\[
\frac{1}{\omega_n+i\omega}\frac{e^{(\omega_n+i\omega)\tau}}{\tau^{3/2}}\left[1+\frac{3}{2(\omega_n+i\omega)\tau}\right]_{t-P}^t.
\]
It is clear that for large values of $\tau$ the second terms in the two brackets are of the order of ${\cal O}(\tau^{-1})$ and therefore, they can be neglected. As a result, using the approximation (\ref{eq64.1}) the integral $R(z,t)$ can be given as as
\begin{equation}
    \label{eqA3}
   R(z,t)\approx\frac{e^{(\omega_n+i\omega)t}}{t^{3/2}(\omega_n+i\omega)}\left[1-e^{-(\omega_n+i\omega)P}\right].
\end{equation}
\end{appendix}

\appendix


\bsp	
\label{lastpage}
\end{document}